\PassOptionsToPackage{table,xcdraw}{xcolor}
\documentclass[journal,twoside,web]{ieeecolor}

\let\rowcolor\relax  

\usepackage{xcolor}
\usepackage{colortbl}

\usepackage{jsen}
\usepackage{cite}
\usepackage{amsmath,amssymb,amsfonts}
\usepackage{graphicx}
\graphicspath{ {images/} }
\usepackage{textcomp}
\usepackage{wrapfig}
\usepackage{bm}
\usepackage{algcompatible}
\usepackage{graphics}
\usepackage{adjustbox}
\let\labelindent\relax
\usepackage{enumitem}
\usepackage{array}
\usepackage{import}
\usepackage{etoolbox}

\makeatletter
\let\@latex@warning@no@line\@gobble
\@ifpackageloaded{xcolor}{%
  \definecolor{black}{rgb}{0,0,0}
  \definecolor{white}{rgb}{1,1,1}
}{}
\makeatother

\def\BibTeX{{\rm B\kern-.05em{\sc i\kern-.025em b}\kern-.08em
    T\kern-.1667em\lower.7ex\hbox{E}\kern-.125emX}}
    
\definecolor{abstractbg}{rgb}{0.89804,0.94510,0.83137}
\begin{document}
\title{Raw Audio Classification with Cosine Convolutional Neural Network (CosCovNN)}
\author{Kazi Nazmul Haque, Rajib Rana, Tasnim Jarin and Bj\"orn W.\ Schuller, Jr., \IEEEmembership{Fellow, IEEE}
\thanks{Date of publication xxxx 00, 0000, date of current version xxxx 00, 0000.}
\thanks{Digital Object Identifier XXXXXX/ACCESS.XXXX.DOI}
\thanks{Manuscript submitted (date) This work was supported in part by the University of Southern Queensland, Australia.}
\thanks{K. N. Haque is with the University of Southern Queensland, Australia (e-mail: shezan.huq@gmail.com).}
\thanks{R. Rana is with the University of Southern Queensland, Australia.(e-mail: rajib.rana@usq.edu.au)}
\thanks{T. Jarin is with the University of Jahangirnagar University, Bangladesh.(tasnim.jarin.2000@gmail.com)}
\thanks{B. W. Schuller, Jr. is with GLAM -- Group on Language, Audio, \& Music, Imperial College London, UK, and also with the Chair of Embedded Intelligence for Health Care and Wellbeing, University of Augsburg, Germany.}}

\IEEEtitleabstractindextext{%
\fcolorbox{abstractbg}{abstractbg}{%
\begin{minipage}{\textwidth}%

\begin{abstract}
This study explores the field of audio classification from raw waveform using Convolutional Neural Networks (CNNs), a method that eliminates the need for extracting specialised features in the pre-processing step. Unlike recent trends in literature, which often focuses on designing frontends or filters for only the initial layers of CNNs, our research introduces the Cosine Convolutional Neural Network (CosCovNN) replacing the traditional CNN filters with Cosine filters. The CosCovNN surpasses the accuracy of the equivalent CNN architectures with approximately $77\%$ less parameters. Our research further progresses with the development of an augmented CosCovNN named Vector Quantised Cosine Convolutional Neural Network with Memory (VQCCM), incorporating a memory and vector quantisation layer. VQCCM achieves state-of-the-art (SOTA) performance across five different datasets in comparison with existing literature. Our findings show that cosine filters can greatly improve the efficiency and accuracy of CNNs in raw audio classification.
\end{abstract}

\end{minipage}}}

\maketitle

\section{Introduction}
\label{sec:introduction}
Convolutional Neural Networks have achieved remarkable success in computer vision, mainly due to their capability to model directly from raw data, thereby removing the need for handcrafted features \cite{jiao2019survey}. Similarly, in the field of audio classification, there is a growing interest in direct raw waveform modelling, which presents unique challenges. The high dimensionality and complex temporal dependencies in audio data require advanced, computationally robust CNN architectures \cite{kim2019comparison}. Directly modelling from raw waveforms eliminates the need for extensive pre-processing, aligning with deep learning's data-driven approach \cite{sangurbanraw}. Unlike traditional spectrogram-based CNNs, limited to specific frequencies, CNNs processing raw waveforms can detect a broader range of frequency responses, increasing their effectiveness as data availability grows \cite{jung2019rawnet}.

In response to these advancements, researchers have explored various modifications of CNN architectures to handle raw audio waveforms more effectively \cite{tuske2014acoustic, Liang2020, 7952651, 7952190}. Much of this work has centred on designing front-end modules and filters, significantly enhancing the initial layers of waveform-based CNNs. This focus stems from the idea that the initial layer plays a crucial role by directly interacting with the raw data. Notable contributions include the SincNet filter by Ravanelli et al. \cite{ravanelli2018speaker}, tailored for initial CNN layers, and Zeghidour et al.'s LEAF \cite{zeghidour2021leaf}, a flexible front-end that adapts to various neural networks. These advances shift away from handcrafted features, instead introducing learnable filters or front-ends in early processing stages, though they still rely on traditional CNN architecture in later layers.

Expanding on this direction, we propose a novel approach: CosCovNN (Cosine Convolutional Neural Network), which incorporates a cosine filter into the CNN framework. This filter, replacing traditional CNN kernels, draws inspiration from the Discrete Cosine Transform (DCT) and the real parts of the Fourier Transform. The DCT is widely recognised for representing signals through a summation of cosine functions at varying frequencies, effectively capturing audio signals' spectral characteristics \cite{1672377, 8683636, geng2020end}. Additionally, the Fourier Transform's real parts, composed of cosine elements, emphasise symmetrical components within a signal's frequency range—fundamental in audio processing, particularly in identifying rhythmic and harmonic structures \cite{smith2008mathematics, 4014108, 8063868}. Our cosine-based filter choice aligns with traditional signal processing principles, enhancing CNN's ability to interpret complex audio patterns.

To further extend CosCovNN's capabilities, we introduce a Vector Quantisation layer immediately following the first convolutional layer, guiding the model to focus on extracting significant features from the audio waveform \cite{8702105}. Additionally, a memory module allows important information captured by the initial layers to be effectively retained and passed through to later stages, improving raw audio modelling performance \cite{Park_Kim_Lee_Bae_Yoon_2018, pmlr-v48-santoro16, graves2014neural}.

In this paper, we make the following key contributions:

\begin{itemize}
\item We introduce CosCovNN, a novel CNN architecture with learnable cosine filters that outperforms traditional CNN models while reducing parameters by approximately 77\%.
\item We present the Vector Quantised Cosine Convolutional Neural Network with Memory (VQCCM), an advanced iteration of CosCovNN, which combines vector quantisation and memory mechanisms. Extensive evaluations of CosCovNN and VQCCM on five distinct datasets highlight VQCCM's state-of-the-art performance, setting new benchmarks in several cases. This underscores the efficacy and potential of our cosine filter-based CNN models for raw audio classification.
\end{itemize}

\section{BACKGROUND AND RELATED WORK}
\label{sec:background}
The field of audio classification has been fundamentally transformed by deep learning, transitioning from traditional signal processing techniques—such as Support Vector Machines (SVM), K-Nearest Neighbors (KNN), and Hidden Markov Models (HMM) \cite{guo2003content, chen2006mixed, lu2002content}—to advanced neural network architectures \cite{mitra2008content, freeman2007audio, shao2003applying, AKKEM2024107881, AKKEM2023105899}. Early models relied heavily on hand-crafted features, but with the introduction of deep learning, particularly Convolutional Neural Networks (CNNs) \cite{khamparia2019sound, hershey2017cnn, yang2019music} and Recurrent Neural Networks (RNNs) \cite{makropoulos2023convolutional, makino2019recurrent, zhang2021attention}, audio classification entered a new era. This evolution was driven by the need to capture complex, high-dimensional patterns in audio data, for which deep learning models are well-suited due to their capacity to learn hierarchical representations directly from raw data \cite{zaman2023survey}.

Despite these advancements, a substantial number of deep learning models for audio classification still rely on two-dimensional (2D) representations of audio data, such as Mel-Frequency Cepstral Coefficients (MFCC), filter banks (FBANK), and spectrograms \cite{arias2021multi, vakkantula2020speech, variani2014deep, heigold2016end}. While effective, these representations are rooted in hand-crafted processing and may not fully harness the learning potential of deep models. Ravanelli and Bengio \cite{ravanelli2018speaker} argued that such features, while perceptually motivated, may not be optimal for all tasks. To address this, they introduced SincNet, a model that learns band-pass filters directly from the raw waveform, allowing for efficient feature learning and faster convergence. SincNet’s approach, with parameterised sine functions defining filter properties, represents a broader trend toward learnable front-ends capable of working directly on raw waveforms \cite{kim2019comparison}. Loweimi et al. extended this idea by developing more flexible kernel-based filters, such as triangular, gammatone, and Gaussian filters, which offer improved alignment with perceptual audio characteristics compared to SincNet’s rectangular filters \cite{loweimi2019learning}.

Furthering this trend, Noé et al. \cite{noe2020cgcnn} introduced CGCNN, which integrates complex Gabor filters and manages the resulting complex-valued signals using Complex-Valued CNNs (CVCNNs). This approach leverages the strong time-frequency localisation properties of Gabor filters, tailoring them for specialised audio applications. Additionally, Sainath et al. \cite{noe2020cgcnn} proposed the Convolutional Long Short-Term Memory Deep Neural Network (CLDNN), demonstrating that raw waveform data, when processed with a sophisticated CLDNN architecture, could match the performance of traditional filter bank models. CLDNN’s time convolution layer effectively captures temporal variations, while the LSTM layers enhance temporal modelling—a significant advance for raw waveform processing.

In response to the limitations of mel-filterbank-based models, Zeghidour et al. introduced time-domain filterbanks (TD-filterbanks) \cite{zeghidour2018learning}, which operate directly on raw audio. Their follow-up work, LEAF, established a fully learnable front-end that dissects mel-filterbanks into functional components—filtering, pooling, compression, and normalisation—creating a versatile architecture that excels across various audio classification tasks, including speech, music, and environmental sounds \cite{zeghidour2021leaf}. EfficientLEAF, proposed by Schlüter and Gutenbrunner \cite{schluter2022efficientleaf}, further reduces computational costs, optimising LEAF’s performance for long input sequences. However, neither LEAF nor EfficientLEAF consistently outperforms traditional mel-filterbanks in all scenarios, indicating that the search for an ideal learnable front-end is ongoing.

Aligned with deep learning’s principle of minimising preprocessing, an increasing area of research is shifting away from traditional features or front-ends, focusing instead on feeding raw waveform data directly into neural networks, particularly one-dimensional (1D) CNNs. This approach aligns with the deep learning ethos of minimal preprocessing and self-sufficient feature extraction. Research has introduced several adaptations of CNN architectures to effectively handle 1D audio signals in their raw form \cite{tuske2014acoustic, palaz2015analysis, hoshen2015speech}. Kim et al. \cite{kim2019comparison} explored this direction further with SampleCNN, an end-to-end architecture optimised for raw waveform processing. SampleCNN’s use of small filter sizes allows it to effectively capture nuances in tasks like music tagging, keyword spotting, and acoustic scene classification, demonstrating the capability of CNNs to work with raw audio directly.

Advancing the direct modelling of raw audio data in CNN architectures, this study introduces CosCovNN, a CNN model that uses cosine-based filters rather than traditional kernels. Inspired by the periodic nature of audio signals and Fourier analysis principles, CosCovNN’s cosine filters provide an efficient and compact representation of audio patterns \cite{salau2019audio}. This choice reduces the model’s complexity by approximately 77

The field of audio classification has been fundamentally transformed by deep learning, shifting from traditional signal processing techniques—such as Support Vector Machines (SVM), K-Nearest Neighbors (KNN), and Hidden Markov Models (HMM) \cite{guo2003content, chen2006mixed, lu2002content}—to advanced neural network architectures \cite{mitra2008content, freeman2007audio, shao2003applying, AKKEM2024107881, AKKEM2023105899}. Early models heavily relied on hand-crafted features; however, the introduction of deep learning, particularly Convolutional Neural Networks (CNNs) \cite{khamparia2019sound, hershey2017cnn, yang2019music} and Recurrent Neural Networks (RNNs) \cite{makropoulos2023convolutional, makino2019recurrent, zhang2021attention}, marked a new era in audio classification. This evolution was driven by the need to capture complex, high-dimensional patterns in audio data, for which deep learning models are well-suited due to their ability to learn hierarchical representations directly from raw data \cite{zaman2023survey}.

Despite these advancements, many deep learning models for audio classification still depend on two-dimensional (2D) representations of audio data, such as Mel-Frequency Cepstral Coefficients (MFCC), filter banks (FBANK), and spectrograms \cite{arias2021multi, vakkantula2020speech, variani2014deep, heigold2016end}. While effective, these representations are based on hand-crafted techniques and may not fully exploit the potential of deep learning models. Ravanelli and Bengio \cite{ravanelli2018speaker} argued that such features, despite being perceptually motivated, may not be optimal for all tasks. To address this, they introduced SincNet, a model that learns band-pass filters directly from the raw waveform, enabling efficient feature learning and faster convergence. SincNet’s approach, which uses parameterized sine functions to define filter properties, exemplifies a broader trend toward learnable front-ends capable of processing raw waveforms directly \cite{kim2019comparison}. Loweimi et al. expanded on this concept by developing more flexible kernel-based filters—such as triangular, gammatone, and Gaussian filters—that offer better alignment with perceptual audio characteristics compared to the rectangular filters used by SincNet \cite{loweimi2019learning}.

Continuing this trend, Noé et al. \cite{noe2020cgcnn} introduced CGCNN, which integrates complex Gabor filters and manages the resulting complex-valued signals using Complex-Valued CNNs (CVCNNs). This approach capitalizes on the strong time-frequency localization properties of Gabor filters, tailoring them for specialized audio applications. Additionally, Sainath et al. \cite{noe2020cgcnn} proposed the Convolutional Long Short-Term Memory Deep Neural Network (CLDNN), demonstrating that when raw waveform data is processed through a sophisticated CLDNN architecture, it can match the performance of traditional filter bank models. CLDNN’s time convolution layer effectively captures temporal variations, while the LSTM layers enhance temporal modeling—representing a significant advance in raw waveform processing.

In response to the limitations of mel-filterbank-based models, Zeghidour et al. introduced time-domain filter banks (TD-filterbanks) \cite{zeghidour2018learning}, which operate directly on raw audio. Their follow-up work, LEAF, established a fully learnable front end that dissects mel-filterbanks into functional components—filtering, pooling, compression, and normalization—creating a versatile architecture that excels across various audio classification tasks, including speech, music, and environmental sounds \cite{zeghidour2021leaf}. EfficientLEAF, proposed by Schlüter and Gutenbrunner \cite{schluter2022efficientleaf}, further reduces computational costs, optimizing LEAF’s performance for long input sequences. However, neither LEAF nor EfficientLEAF consistently outperforms traditional mel-filterbanks in all scenarios, indicating that the search for the ideal learnable front end is ongoing.

Aligned with deep learning's principle of minimizing preprocessing, an increasing area of research is moving away from traditional features or front ends, focusing instead on feeding raw waveform data directly into neural networks, particularly one-dimensional (1D) CNNs. This approach aligns with the deep learning ethos of minimal preprocessing and self-sufficient feature extraction. Research has introduced several adaptations of CNN architectures to effectively handle 1D audio signals in their raw form \cite{tuske2014acoustic, palaz2015analysis, hoshen2015speech}. Kim et al. \cite{kim2019comparison} explored this direction further with SampleCNN, an end-to-end architecture optimized for raw waveform processing. SampleCNN’s use of small filter sizes allows it to effectively capture nuances in tasks such as music tagging, keyword spotting, and acoustic scene classification, demonstrating the capacity of CNNs to work directly with raw audio.

Advancing the direct modeling of raw audio data in CNN architectures, this study introduces CosCovNN, a CNN model that utilizes cosine-based filters designed specifically for audio processing. In Table!\ref{table:comparison}, we contrast CosCovNN with other closely related proposals in the literature, showcasing its uniqueness and novelties.

\begin{table*}[ht]
\centering
\caption{Comparison of VQCCM with the literature}
\begin{adjustbox}{max width=\textwidth}
\LARGE
\begin{tabular}{|>{\raggedright\arraybackslash}p{3cm}|>{\raggedright\arraybackslash}p{5cm}|>{\raggedright\arraybackslash}p{5cm}|>{\raggedright\arraybackslash}p{5cm}|>{\raggedright\arraybackslash}p{5cm}|>{\raggedright\arraybackslash}p{5cm}|>{\raggedright\arraybackslash}p{5cm}|>{\raggedright\arraybackslash}p{5cm}|}
\hline
\textbf{\LARGE Aspect} & \textbf{\LARGE Our Study (VQCCM)} & \textbf{\LARGE TD-filterbanks\cite{}} & \textbf{\LARGE SincNet} & \textbf{\LARGE LEAF} & \textbf{\LARGE SampleCNN} & \textbf{\LARGE CGCNN} & \textbf{\LARGE CLDNN} \\ \hline

\textbf{\LARGE Advantages} & 
\begin{itemize}[leftmargin=*]
    \item \LARGE Achieved state-of-the-art results on five datasets.
    \item \LARGE Low computational complexity due to reduced parameter count with cosine filters.
    \item \LARGE Can be applied to various audio classification tasks.
    \item \LARGE VQ and Memory layers enhance the model's adaptability to different datasets.
\end{itemize} & 
\begin{itemize}[leftmargin=*]
    \item \LARGE Improved performance over traditional MFCC.
    \item \LARGE Learnable filters optimise the feature extraction process.
    \item \LARGE Effective in phone recognition tasks.
    \item \LARGE Learnable filters can adapt to different types of audio signals.
\end{itemize} & 
\begin{itemize}[leftmargin=*]
    \item \LARGE Better performance in speaker recognition tasks.
    \item \LARGE Efficient filter design with parametrised sine functions.
    \item \LARGE Effective in speaker recognition tasks.
    \item \LARGE Parametrised sine functions can adapt to various audio patterns.
\end{itemize} & 
\begin{itemize}[leftmargin=*]
    \item \LARGE Outperforms mel-filterbanks across various tasks.
    \item \LARGE Lightweight architecture with consistent improvements in multi-task settings.
    \item \LARGE General-purpose learnable frontend.
    \item \LARGE Can be applied to speech, music, and environmental sounds.
\end{itemize} & 
\begin{itemize}[leftmargin=*]
    \item \LARGE Effective in various audio classification tasks.
    \item \LARGE Small filter sizes enhance computational efficiency.
    \item \LARGE Effective across different audio classification tasks.
    \item  \LARGE Enhanced with residual and squeeze-and-excitation networks.
\end{itemize} & 
\begin{itemize}[leftmargin=*]
    \item \LARGE Superior time-frequency localisation with Gabor filters.
    \item \LARGE Efficient handling of complex-valued signals.
    \item \LARGE Tailored for specialised audio applications.
    \item  \LARGE Gabor filters can be adapted to different time-frequency requirements.
\end{itemize} & 
\begin{itemize}[leftmargin=*]
    \item \LARGE Raw waveform features matched performance of traditional methods.
    \item \LARGE Effective time convolution and LSTM layers for temporal modelling.
    \item \LARGE Effective in matching performance of traditional methods.
    \item \LARGE Can be applied to various temporal modelling tasks.
\end{itemize} \\ \hline

\textbf{\LARGE Limitations} & 
\begin{itemize}[leftmargin=*]
    \item \LARGE Evaluated on only five datasets.
    \item \LARGE Needs optimisation for real-time processing.
\end{itemize} & 
\begin{itemize}[leftmargin=*]
    \item \LARGE Limited to phone recognition tasks.
    \item \LARGE Higher complexity with convolutional layers.
\end{itemize} & 
\begin{itemize}[leftmargin=*]
    \item \LARGE Primarily evaluated on speaker recognition tasks.
    \item \LARGE Limited training data.
\end{itemize} & 
\begin{itemize}[leftmargin=*]
    \item \LARGE Initial reliance on convolutional architecture.
    \item \LARGE Computationally intensive for large-scale applications.
\end{itemize} & 
\begin{itemize}[leftmargin=*]
    \item \LARGE Primarily focused on small-scale datasets.
    \item \LARGE Computational complexity can be high for Large datasets.
\end{itemize} & 
\begin{itemize}[leftmargin=*]
    \item \LARGE Complexity of managing complex-valued signals.
    \item \LARGE Limited evaluation on diverse tasks.
\end{itemize} & 
\begin{itemize}[leftmargin=*]
    \item \LARGE Requires sophisticated architecture.
    \item \LARGE High computational cost.
\end{itemize} \\ \hline

\textbf{\LARGE Inferences Taken} & 
\begin{itemize}[leftmargin=*]
    \item \LARGE VQ and Memory layers substantially boost model performance.
    \item \LARGE Cosine filters reduce parameter count and enhance efficiency.
\end{itemize} & 
\begin{itemize}[leftmargin=*]
    \item \LARGE Learning filters directly from raw audio can outperform traditional feature extraction methods.
    \item \LARGE Importance of filter initialisation.
\end{itemize} & 
\begin{itemize}[leftmargin=*]
    \item \LARGE Compact filters are effective for speaker recognition.
    \item \LARGE High-level tune-able parameters improve performance.
\end{itemize} & 
\begin{itemize}[leftmargin=*]
    \item \LARGE Learning all operations of audio feature extraction can outperform traditional methods.
    \item \LARGE Important to evaluate across diverse tasks.
\end{itemize} & 
\begin{itemize}[leftmargin=*]
    \item \LARGE Fine-tuned CNN architectures can harness nuances of raw audio data effectively.
\end{itemize} & 
\begin{itemize}[leftmargin=*]
    \item \LARGE Gabor filters provide effective time-frequency localisation.
    \item \LARGE Complex-valued signals can be managed with CVCNNs.
\end{itemize} & 
\begin{itemize}[leftmargin=*]
    \item \LARGE Raw waveform data can be effectively used with sophisticated models.
    \item \LARGE Temporal convolution and modelling are crucial for performance.
\end{itemize} \\ \hline

\end{tabular}
\end{adjustbox}
\label{table:comparison}
\end{table*}

\section{PROPOSED RESEARCH METHOD}
\label{sec:proposed_method}
\begin{figure*}[!t]
\centerline{\includegraphics[width=\linewidth]{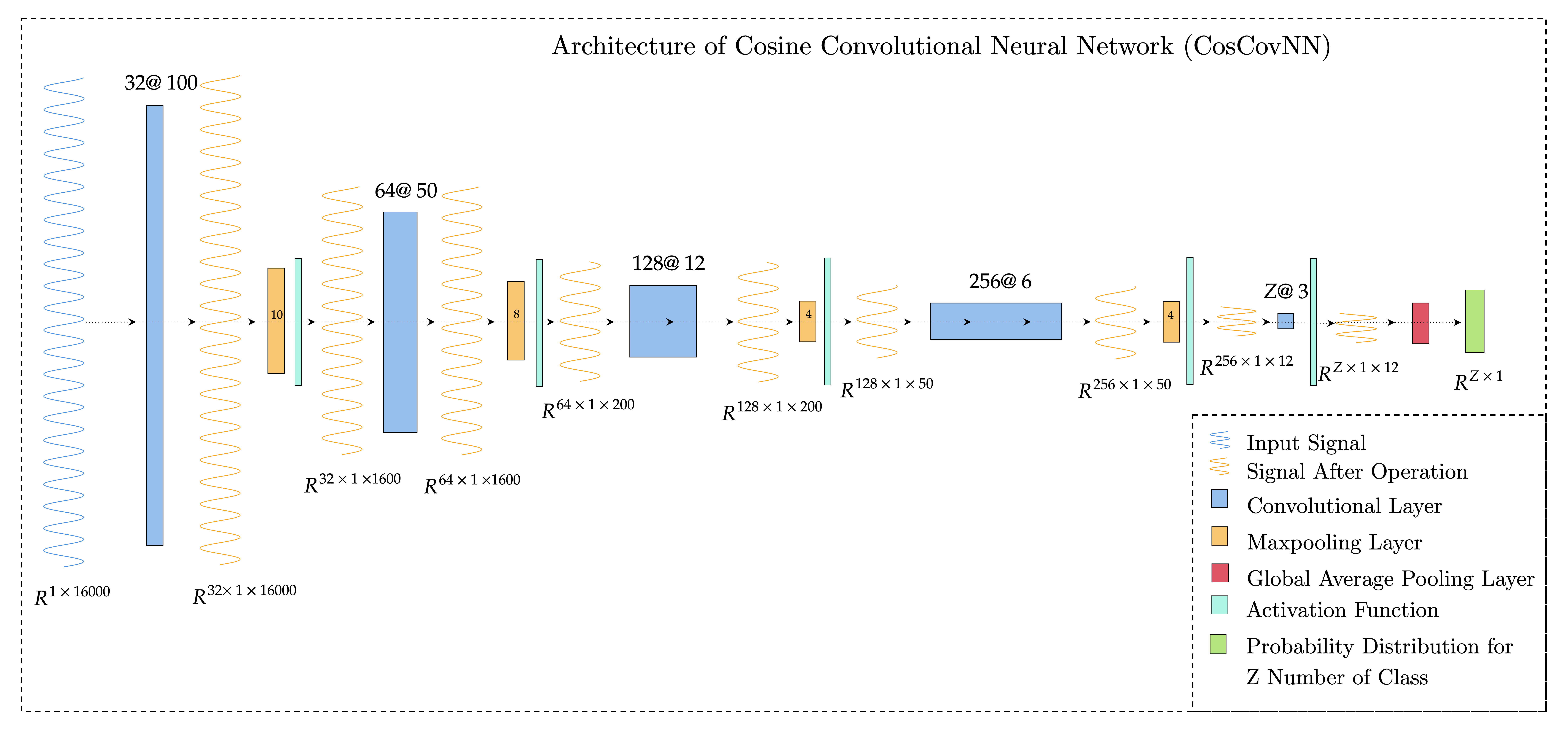}}
\caption{The present figure depicts an exposition of the intricate architecture of the CosCovNN model, which is designed to process a one-second audio signal with a sampling rate of 16 kHz. The convolutional layer is illustrated with the number of filters denoted by the symbol `@` on the left-hand side, while the filter size is indicated on the right-hand side.}
\label{fig:cnn_vs_coscov}
\end{figure*}

\subsection{Background Knowledge}
This section presents a detailed overview of the Cosine Convolutional Neural Network  architecture and its integration into the Vector Quantised Cosine Convolutional Neural Network with Memory (VQCCM) model. The VQCCM model is constructed using the CosCovNN layer as a fundamental building block.

\subsubsection{Convolution in Convolutional Neural Network}
1D convolutional neural networks usually consist of convolutional layers, maxpool, and fully connected layers. For any particular layer, if the input is $x[n]$, the convolutional operation can be defined as follows,

\begin{equation}
\label{eq:convolution}
\begin{aligned}
o[n] = x[n] * h[n] = \sum\limits_{l=0}^{L-1}{x[l] \cdot h[n-l]}\\
\end{aligned}
\end{equation}

where, $h[n]$ is the filter with length $L$ and $o[n]$ is the output of that layer. During training, the aim is to learn the $L$ parameters of the filters. Each layer of the convolutional neural network is comprised of multiple filters. We learn these filters through back-propagation from the training data during the training.  

\subsubsection{Vector Quantisation}

The idea behind vector quantisation ($VQ$) is to represent $n$ set of vectors, $V \in \{v_{1}, v_{2}, \dots v_{n}\}$ by a finite set of $m$ representative vectors from a codebook, $C \in \{c_{1}, c_{2}, \dots c_{m}\}$. Here, each vector $v_{i}$ and $c_{j}$ has an equal dimension of $D$ where $ i \in \{1,2,3 \dots n\}$ and $j \in \{1,2,3 \dots m\}$ \cite{van2017neural}. The goal of $VQ$ is to find the closest representative vector of $v_{i}$ in $C$ and represent $v_{i}$ as $c_{j}$ through the mapping function $G$, which can be formulated as follows,
\begin{equation}
\label{eq:vq_distance}
\begin{aligned}
G(v_{i}) = argmin_{j} ||v_i - c_j||_{2} 
\end{aligned}
\end{equation}
where $||v_i - c_j||_{2} $ represents the squared Euclidean distance between the input vector point $v_i$ and the representative vector $c_j$.

$VQ$ representation can be a very powerful layer in neural networks; however, the challenge lies in the computation of the gradient for $argmin_{j} ||v_i - c_j||_{2}$. In the VQ-VAE paper \cite{van2017neural}, authors have addressed this issue by using the gradient of $\nabla_{c_{j}}L$ to update the vector $v_i$, where $L$ is the loss of any neural network. In this paper, the authors have used the $VQ$ layer in their Autoencoder Network to learn the discrete representation $c_{j}$ for $v_{i} = E (x_{i})$, where $E$ is the Encoder, and $x_i$ is the input data. Decoder, $D$ takes the $VQ$ representation $c_{j}$ to reconstruct $x_i$. Here, the reconstruction objective is $ \log D(\hat{x_{i}}\approx x_{i} | c_{j})$. As the dimension of $c_{j}$ is equal to $v_{i}$, the gradient calculated for the $c_{j}$ can be used to update the weights of $E$. This way, the Autoencoder is trained end-to-end with the back-propagation algorithm. Here, the authors have the following loss function to train the VQ-VAE,

\begin{equation}
\label{eq:vq_vae}
\begin{aligned}
L = \log D(\hat{x_{i}} \approx x_{i}| c_{j}) + \|\text{sg}[v_{i}] - c_{j}\|^2_2 + \beta \|v_{i} - \text{sg}[c_{j}]\|^2_2,
\end{aligned}
\end{equation}

Where, stop gradient, $sg$ stops the flow of gradient during the back-propagation through a particular layer in a neural network and $\beta$ is the hyperparameter. Here, $\beta \|v_{i} - \text{sg}[c_{j}]\|^2_2$  part forces the Encoder, $E$ to learn $v_{i}$ close to $c_{j}$ and $\|\text{sg}[v_{i}] - c_{j}\|^2_2$  part makes sure that $c_{j}$ does not deviate much from $v_{i}$.

\subsection{Architecture of the CosCovNN}
\begin{figure*}[!t]
\centerline{\includegraphics[width=\linewidth]{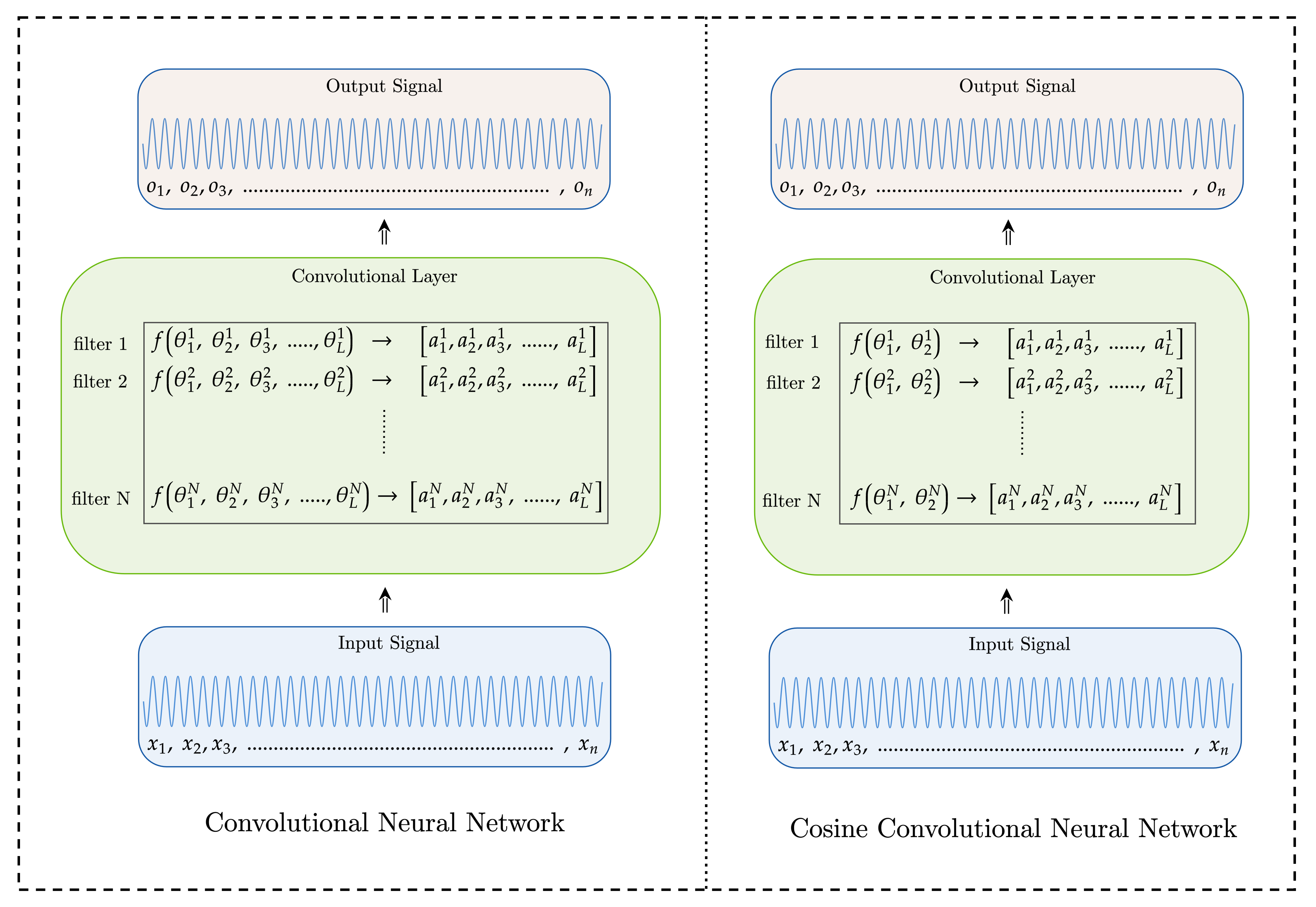}}
\caption{The figure illustrates a comparative analysis between the convolutional layer of a standard convolutional neural network (CNN) and that of the Cosine Convolutional Neural Network (CosCovNN). The symbols $f$, $x$, $o$, $\theta$, and $a$ denote the filter, input, output, learnable parameters, and filter values, respectively, for both models. Notably, the CosCovNN architecture has only two parameters, $\theta$, for any given filter of size $L$, whereas the CNN model requires $L$ parameters for each filter of size L}
\label{fig:cnn_coscov_filter}
\end{figure*}

The 1D Convolutional Neural Network (CNN) and 1D Cosine Convolutional Neural Network (CosCovNN) differ primarily in the filter of their convolutional layers. A CNN requires learning $L$ parameters for each particular filter of size $L$, whereas a CosCovNN only requires learning two parameters for a filter of the same size $L$. Thus, for any given filter, the CosCovNN effectively reduces the number of parameters by $L-2$, where $L > 2$, relative to the CNN architecture. A visual comparison of the convolutional layers of the two architectures is presented in Fig. \ref{fig:cnn_coscov_filter}. Important components of the architecture are discussed as follows,

\subsubsection{Convolutional Layer}
For the convolutional layer, we generate the filter from a periodic cosine function. A cosine function can be represented as follows,

\begin{equation}
\label{eq:periodicfunction}
\begin{aligned}
y[n] = A \cos (\frac{2\pi}{\lambda}n  )\\
\end{aligned}
\end{equation}

where, A is the Amplitude, $\lambda$ is the wave length and $n$ is the step. As, $2\pi$ is a constant,  let  $\frac{2\pi}{\lambda} = \theta_{2}$ and $A = \theta_{1}$. Therefore, we can represent the equation as follows,
\begin{equation}
\label{eq:filterfunction}
\begin{aligned}
g[n,\theta_{1},\theta_{2}] = \theta_{1} \cos (\theta_{2}n )\\
\end{aligned}
\end{equation}

Here, for any particular convolutional layer of the CosCov-Net, if the input is $x[n]$, the convolutional operation can be defined as follows,

\begin{equation}
\label{eq:convolutionCosCov}
\begin{aligned}
o[n] = x[n] * g[n,\theta_{1},\theta_{2}] \\
\end{aligned}
\end{equation}
Here, $o[n]$ is the output, $\theta_{1}$, $\theta_{2}$ are the learn-able parameters of the filter.

\subsubsection{Pooling Layer}

We have used 1D max-pooling layer for down-sampling between layers. This layer selects the most salient features within a window of size $k$, thereby enhancing the significance of the features obtained from the convolutional layer.\cite{kiranyaz20211d}

\subsubsection{Activation Layer}

The activation function employed in CosCovNN is a crucial component of the network's architecture. As the values of the filters in CosCovNN are periodic, ranging from $-1$ to $1$, it is imperative to maintain this range throughout the output of each layer of the network. To ensure consistency in the range of output values, we utilise the $tanh$ activation function.

\subsubsection{Classification Layer}

In classification tasks involving $Z$ classes, the conventional approach is to employ a fully connected layer of size $Z$ at the end of the network. However, this results in a substantial increase in the number of model parameters. To address this issue, we propose utilising $Z$ Cosine Convolutional Layers with global average pooling \cite{lin2013network} in the classification layer, which enables us to significantly reduce the parameter count.

\subsubsection{Dropout}

To enhance the resilience of the network and prevent overfitting, we incorporate 1D spatial dropout \cite{tompson2015efficient}. Unlike conventional dropout methods that randomly discard individual elements, spatial dropout removes entire 1D feature maps, thereby enabling the network to learn more robust and generalised features.

\subsection{Architecture of the VQCCM}

\begin{figure*}[!t]
\centerline{\includegraphics[width=\linewidth]{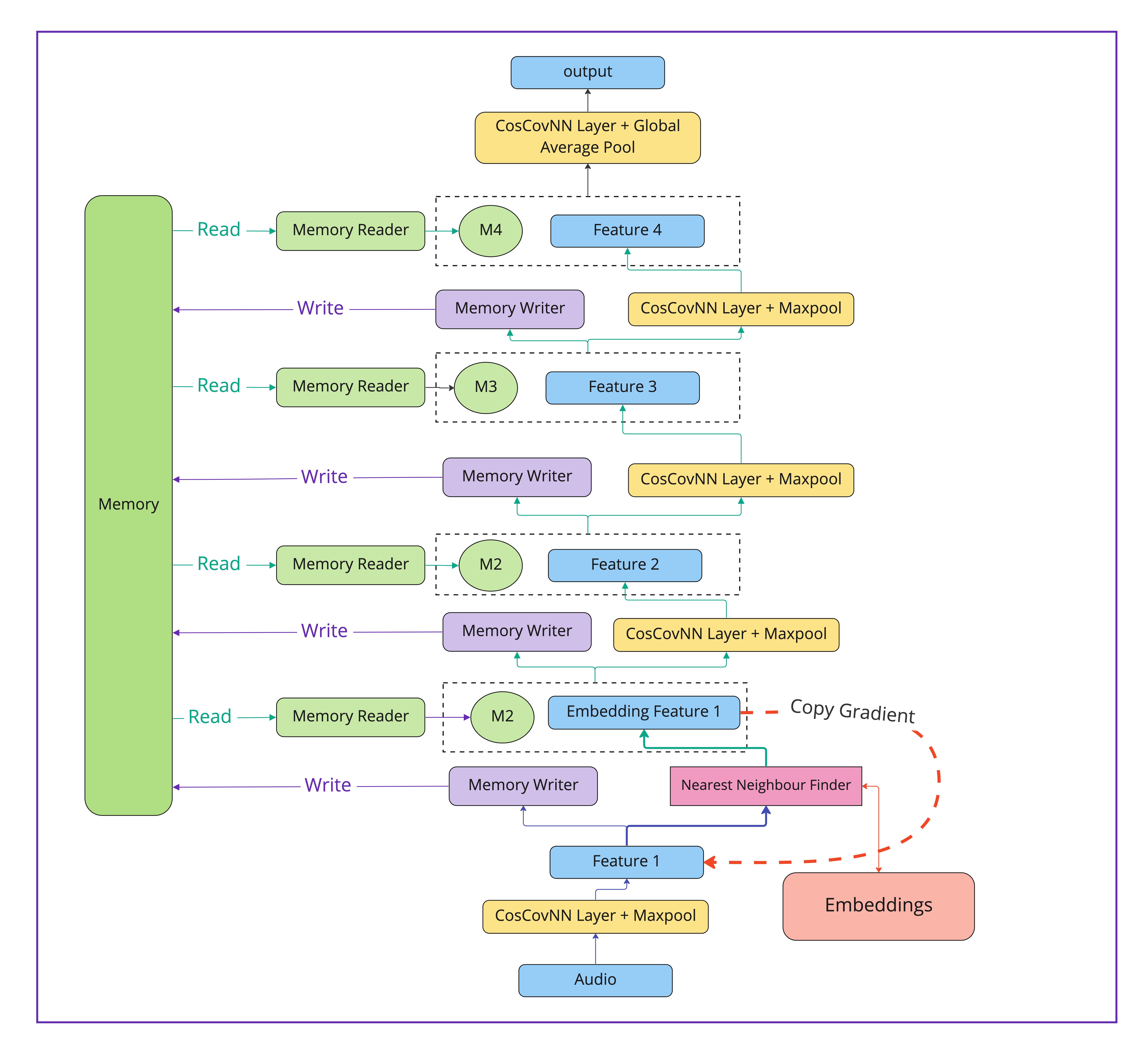}}
\caption{The diagram presents the structure of the VQCCM model. In this architecture, the Vector Quantisation Layer is strategically positioned right after the initial CosCovNN layer. This placement compels the model to transmit only crucial information to subsequent layers, thereby ensuring that the first layer extracts significant features due to its direct interaction with the input. On the left side of the illustration, the memory layer is depicted. This layer is connected to every other layer in the model, facilitating the transfer of information throughout the entire network, from the initial layers to the final ones.}
\label{fig:vq_coscov}
\end{figure*}

The VQCCM model is an extension of the CosCovNN that incorporates Vector Quantization (VQ) and Memory Layers to improve its performance. Fig. \ref{fig:vq_coscov} illustrates the detailed architecture of the VQCCM model. In this model, the VQ layer is used in the first layer, and every layer has a memory writer and reader. During training, we learn the memory layer, as well as the reader and writer.

In the first layer, only a memory writer is present, which takes the feature from the preceding layer and writes important information to carry it to the next layer. The memory readers read the information from memory and merge it with the feature. The memory writer writes information in the memory from each layer on top of the memory obtained from the preceding layer. This allows the important information to pass from direct audio input to the output layer and in between.

\subsubsection{Vector Quantisation Layer}

\begin{figure*}[!t]
\centerline{\includegraphics[width=\linewidth]{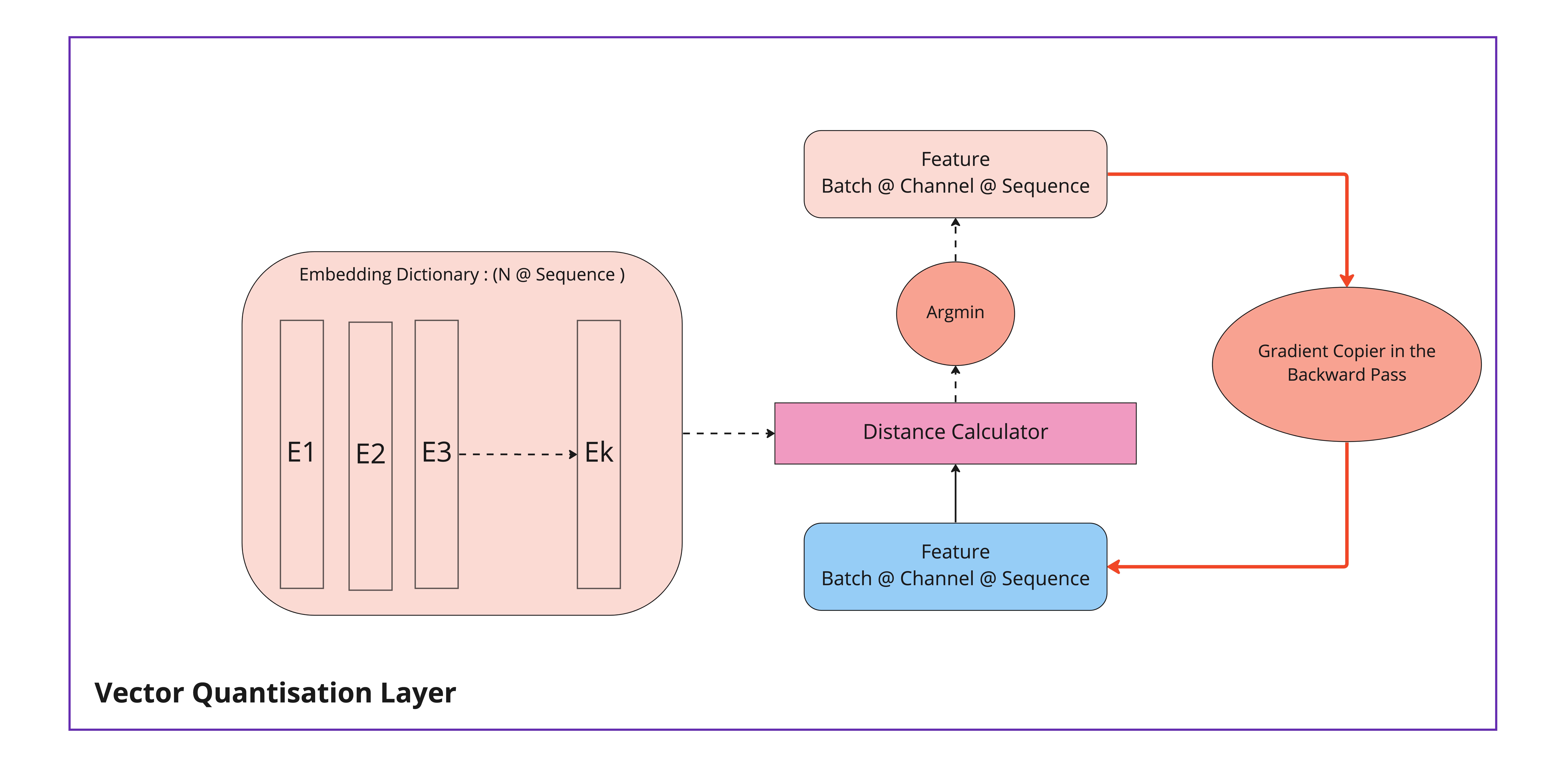}}
\caption{The architectural details of the VQ layer are shown in this figure. Here, the incoming feature is replaced with the nearest embedding from the codebook $E$. This replacement operation with $argmin$ does not have any gradient. Therefore, the gradient is copied from the replaced feature to the original feature.}
\label{fig:vq}
\end{figure*}

In the VQ layer, an embedding matrix or codebook $E \in \mathbb{R}^{d, k}$ is used, where $k$ represents the number of embedding vectors and $d$ is the sequence length of the vector. It is noteworthy that $d$ is identical to the sequence length of the incoming feature.

During the forward pass, the audio is passed through the Cosine Convolutional Neural Network (CosCovNN) and max-pooling layer, yielding a feature representation denoted as $F \in \mathbb{R}^{b, c, d}$, where $b$ and $c$ denote the number of batches and channels, respectively. Specifically, for each batch and channel, the Euclidean Distance between the feature, $F_{i}$, where $ i \in \{1,2, \dots b \times c\}$ and the embedding vectors $ E_{j} $, where $ j \in \{ 1,2, \dots k \} $  from the codebook $E$, is computed. The closest vector ${F_{i}}^{'}$ is then selected from the codebook, 
replacing the original $ F_{i} $ feature  and passing it to the next layer of the model. This operation can be expressed as follows, 

\begin{equation}
\label{eq:vq_layer_vqccm}
\begin{aligned}
F_{i} = F_{i}^{'} = E_{k}, \text{where},  k = argmin_{j} ||F_i - E_j||_{2} 
\end{aligned}
\end{equation}

During the backward pass, the gradient of ${F_{i}}^{'}$ is copied to $F_{i}$ as their shapes are equal, and the training process continues as usual. Fig. \ref{fig:vq} shows the architecture of the VQ layer.

\subsubsection{Memory Layer}

\begin{figure*}[!t]
\centerline{\includegraphics[width=\linewidth]{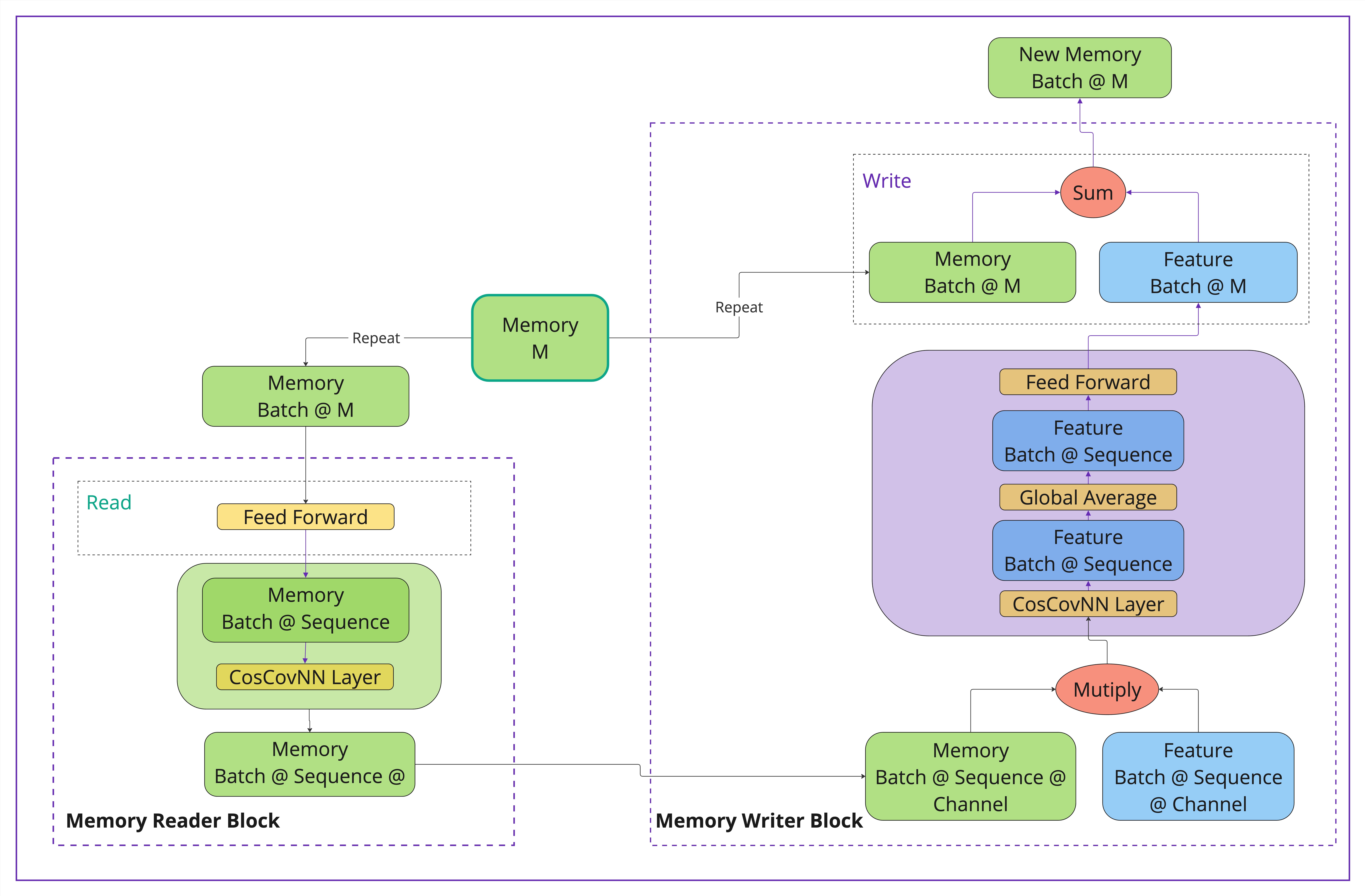}}
\caption{The figure provides a detailed view of the Memory Reader and Writer components within the VQCCM architecture. The Memory, labeled as $M$, is situated centrally between the two blocks. The Reader block accesses the memory, integrating it through a series of feed forward networks and a CosCovNN layer. This processed information is then relayed to the Writer block, where it merges the existing memory with the current layer's features, subsequently generating a new memory state. This updated memory state supersedes the previous one and is passed on to the subsequent layer for further processing }
\label{fig:memory}
\end{figure*}
  
The memory layer is composed of three key components: Memory ($MEM$), Memory Writer ($MW$), and Reader ($MR$). The Memory vector $MEM$ is a vector with dimensions $\mathbb{R}^{1 \times M}$, where $M$ represents the size of the Memory vector.
During each layer, the Reader, $MR$ reads the Memory vector $MEM$ and multiplies it by the current layers feature vector $f_{l}$ to get $f_{l}^{'}$, where $l$ refers to the index of the layer of the VQCCM.
Now, the Memory Writer, $MW$ takes the feature vector $f_{l}^{'}$ and sums its intermediate representation with $MEM$ to produce a new Memory vector, which is utilised by the subsequent layers Reading operation.
By utilising these distinct components, the memory layer facilitates the efficient flow of information across the VQCCM network. Rather than initiating the memory with zero, we learn the $MEM$ during the training and used this learned memory during the test time as the start memory. Fig. \ref{fig:memory} portrays the architecture of the Memory reader and writer blocks.

\paragraph{Memory Reader}

The $MR$ takes the memory, $MEM \in \mathbb{R}^{B,M}$ repeated over the batch of size $B$. The initial read operation is conducted by a Feed Forward Network, $F_{R}$. The output size of $F$ is equal to the sequence length, $S$ of the current layer feature $f_{l} \in \mathbb{R}^{B, S, C}$, where $C$ is the number of channels. Then the output is passed through $C$ number of CosCovNN layers (Cosine Convolutional Layer, $CCL$) to get the memory, $MEM \in \mathbb{R}^ {B,S,C}$. Now, the $MEM$ is multiplied with the feature $f_{l}$ to get the feature $f_{l}^{'}$ to pass through the next layer and the Memory Writer. After both $F_{R}$ and $CCL$ layers, we use the activation function $tanh$. The whole read operation can be summarised as follows,

\begin{equation}
\label{eq:MR1}
\begin{aligned}
f_{l}^{'} =  f_{l} \odot tanh (CCL (tanh (F_{R} (MEM))))
\end{aligned}
\end{equation}

\paragraph{Memory Writer}

The Memory Writer, $MW$ takes the feature $f_{l}^{'}$ and passes it through the $CCL$ layer. Then the output is passed through Global Average Pooling, $GAP$ to remove the dimension $C$ from $f_{l}^{'}$. Now, this is passed through the Feed Forward Network $F_{W}$ to get the intermediate feature of size $( B, M )$. Finally, to write and create a new memory, the intermediate feature is added with the memory, $MEM$. This $MEM$ is used in the subsequent layers read operation. Similar to $MR$, we use $tanh$ activation after each layer. The whole write operation can be expressed as follows,

\begin{equation}
\label{eq:MR}
\begin{aligned}
MEM =  MEM + tanh( F_{W}( GAP( tanh( CCL (f_{l}^{'})))))
\end{aligned}
\end{equation}

\subsubsection{Training Objective}

As both of the networks are evaluated on classification tasks, we used cross-entropy loss during the training. However, for VQCCM, we have an extra loss for the VQ layer. The total loss, $L$ is computed as follows,

\begin{equation}
\label{eq:vqccm_loss}
\begin{aligned}
L = -\sum_{i=1}^Z {y_{i}\log( \hat{y_{i}})} + \|\text{sg}[F_{i}] - E_{j}\|^2_2 + \beta \|F_{i} \- \text{sg}[E_{j}]\|^2_2
\end{aligned}
\end{equation}

Where $Z$ is the number of classes in the classification task.

\section{DATASET}
\label{sec:DATASET}
CosCovNN and VQCCM are evaluated on five datasets from both speech and non-speech audio domains. The datasets used in this study are as follows:

\subsection{Speech Command Classification}
Speech Command Dataset is  consisted of 105,829 utterances of length one second and there are 35 words from 2,618 speakers \cite{Pete_03209}.
An audio digit classification dataset named S09, is created from this speech command dataset where it consists of utterances for different digit categories from zero to nine. Specifically, we have used this dataset for expensive experiments.

\subsection{Speech Emotion Classification}

In our study, we utilised the IEMOCAP dataset for emotion classification. This dataset comprises a total of 12 hours of audio data, consisting of five sessions featuring two distinct speakers (one male and one female) for each session. To ensure consistency with prior research, we focused on four primary emotional states - namely angry, neutral, sad, and happy (with the excitement category consolidated with happy) \cite{iemocap}.

\subsection{Speaker Identification}
We employed the VoxCeleb dataset \cite{voxceleb} for the task of speaker identification/classification. This dataset consists of over 100,000 utterances (1000 hours of audio recordings) from 1251 speakers.

\subsection{Acoustic Scenes Classification}

We have chosen the TUT Urban Acoustic Scenes 2018 dataset for our acoustic scenes classification task. The dataset comprises 24 hours of audio, which is divided into 8640 segments of 10 seconds each. The audio belongs to ten different classes, including 'Airport', 'Shopping mall', 'Metro station', 'Pedestrian street', 'Street with medium level of traffic', 'Travelling by a tram', 'Travelling by a bus', 'Travelling by an underground metro', and 'Urban park'. \cite{mesaros2018multi}

\subsection{Musical Instrument Classification}

We utilised The Nsynth audio dataset to evaluate our models for the musical instrument classification task. This dataset comprises of 305,979 musical notes with a duration of four seconds, representing ten different instruments such as 'brass', 'flute', 'keyboard', 'guitar', 'mallet', 'organ', 'reed', 'string', and 'synth lead', along with one vocal class \cite{nsynth2017}.

\section{EXPERIMENTAL SETUP, RESULTS AND DISCUSSION}

We performed experiments to assess the effectiveness of both CosCovNN and VQCCM. Our evaluation of CosCovNN involved identifying an appropriate baseline architecture to compare its performance with similar CNN architectures. Additionally, we established an experimental setup to analyse the performance of VQCCM relative to state-of-the-art literature.

\subsection{Model Architecture Search for CosCovNN}
To evaluate the performance of cosine convolutional filters, we need to first find a benchmark architecture for the model. This will allow us to evaluate its performance and computational complexity with similar CNN architecture.

\subsubsection{Experimental Setup}
Like any typical CNN model, finding a suitable architecture for CosCovNN is the most challenging part. It is very common to tune the architecture (eg. change the size of kernels, number of layers, and number of filters) of any CNN model according to the datasets. However, tuning the proposed CosCovNN for different datasets is out of the scope of this research work. Therefore, we want to search for an optimal architecture for any single dataset, then measure its performance on different datasets by changing only the number of filters and layers. To find this architecture, we experiment with the S09 dataset. 

First, we fix a backbone network and then change different settings to find the optimal architecture. For the backbone network (BBN), we choose five layers for the CosCovNN with filter size 12 and incremental number of filters as 32, 64, 128, 256 and 10 (number of class for S09 is 10) respectively from layer 1 to 5. For the activation function we use tanh function and maxpool at each layer of size 2 where Dropout is 50 percent.

To find suitable filter size for each layer, we follow the following strategy,
\begin{itemize}
\item Step 1: choose layer 1 from the BBN.
\item Step 2: get classification accuracy changing the size of the filter $F$,  to 3, 6, 12, 25, 50, 100, 200, 300 (while rest of the layers are unchanged in the BBN). 
\item Step 3: choose the filter $F_{best}$, with highest accuracy and replace the filter $F$ of the layer with $F_{best}$ (while rest of the layers are unchanged in the BBN).
\item Step 4: choose next layer from the BBN.
\item Step 5: Go back to step 2 if this not the last layer.
\end{itemize}

After we find the optimal filter size for each of the layers, we follow a similar strategy for the max-pooling layer. We explored 2, 4, 6, 8, 10, 20 window sizes for each layer. Here, every experiment is conducted five times and the maximum accuracy is recorded for comparison.  This approach allowed us to identify the maximum accuracy achievable for any given setting.

\subsubsection{Results}

The results presented in Tables. \ref{tab:class_s09_filter_coscov}, we observed that a gradual reduction in filter size from layer 1 to 5 yielded better performance. This is likely due to a decrease in feature size resulting from maxpooling at each layer, which allows for better capturing of key features with smaller filter sizes.

For maxpooling window sizes, we identified optimal values of 10, 8, 4, and 4 for layers 1 to 4, respectively. Our analysis indicates that larger pooling sizes can lead to significantly improved accuracy. However, balancing pooling sizes across different layers is essential to avoid performance degradation. These results provide valuable insights into optimising the architecture of the CosCovNN model for improved accuracy.

\begin{table}[t!]
\centering
\caption{Classification Accuracy of CosCovNN on the S09 dataset for different filter size}
\label{tab:class_s09_filter_coscov}
\begin{tabular}{|c|c|c|c|c|c|}
\hline
\textbf{Filter Size}  & \textbf{Layer 1} & \textbf{Layer 2} & \textbf{Layer 3} & \textbf{Layer 4} & \textbf{Layer 5} \\ \hline
\textbf{3}   & 78.64            & 79.54            & 83.67            & 84.16            & \textbf{85.22}   \\ \hline
\textbf{6}   & 81.34            & 81.93            & 84.16            & \textbf{84.24}   & 84.52            \\ \hline
\rowcolor[HTML]{FFFFFF} 
\textbf{12}  & 81.89            & 82.12            & \textbf{84.20}   & 84.20            & 84.24            \\ \hline
\textbf{25}  & 82.00            & 83.54            & 84.12            & 83.50            & 84.05            \\ \hline
\textbf{50}  & 82.09            & \textbf{84.20}   & 83.10            & 82.09            & 83.77            \\ \hline
\textbf{100} & \textbf{82.12}   & 83.97            & 81.97            & 81.77            & 83.65            \\ \hline
\textbf{200} & 82.05            & 83.14            & 81.85            & 81.62            & 83.58            \\ \hline
\textbf{300} & 82.01            & 81.81            & 81.30            & 81.58            & 83.54            \\ \hline
\end{tabular}

\end{table}

\begin{table}[t!]
\centering
\caption{Classification Accuracy of CosCovNN on the S09 dataset for different window size of the Maxpool}
\label{tab:class_s09_pool_coscov}
\begin{tabular}{|l|l|l|l|l|}
\hline
\textbf{Pool Size }  & \textbf{Layer 1} & \textbf{Layer 2} & \textbf{Layer 3} & \textbf{Layer 4} \\ \hline
\textbf{2}  & 85.22            & 93.81            & 95.37            & 95.53            \\ \hline
\textbf{4}  & 89.53            & 94.43            & \textbf{95.53}   & \textbf{96.32}            \\ \hline
\textbf{8}  & 92.71            & \textbf{95.37}   & 95.22            & 96.00               \\ \hline
\textbf{10} & \textbf{93.81}   & 92.98               & 94.86            & 95.69            \\ \hline
\textbf{20} & 93.57            & 91.77               & 94.43             & 95.34            \\ \hline
\end{tabular}
\end{table}

\subsection{Comparison between CosCovNN and CNN}

\begin{table*}[t!]
\centering
\caption{Comparison of CosCovNN and VQCCM with the literature based on the test classification accuracy for different tasks}
\label{tab:cl_lit}
\begin{tabular}{|l|l|l|l|l|l|l|}
\hline
\textbf{Classification Task} & \textbf{TD-fbanks} & \textbf{SincNet} & \textbf{CNN} & \textbf{CosCovNN} & \textbf{LEAF} & \textbf{VQCCM} \\ \hline
\textbf{Speech Command} & 87.3 $\pm$ 0.4 & 89.2 $\pm$ 0.4 & 83.1 $\pm$ 0.5 & 91.5 $\pm$ 0.2 & 93.4 $\pm$ 0.3 & \textbf{95.6 $\pm$ 0.1} \\ \hline
\textbf{Spoken Digit} & 94.6 $\pm$ 0.3 & 95.4 $\pm$ 0.2 & 91.4 $\pm$ 0.2 & 96.3 $\pm$ 0.1 & 96.7 $\pm$ 0.2 & \textbf{97.1 $\pm$ 0.2} \\ \hline
\textbf{Speech Emotion} & 58.7 $\pm$ 1.4 & 59.5 $\pm$ 3.2 & 54.2 $\pm$ 1.2 & 63.1 $\pm$ 2.8 & 66.8 $\pm$ 1.8 & \textbf{71.2 $\pm$ 1.4} \\ \hline
\textbf{Acoustic Scenes} & \textbf{99.5 $\pm$ 0.4} & 96.7 $\pm$ 0.9 & 95.6 $\pm$ 0.7 & 98.3 $\pm$ 0.6 & 99.1 $\pm$ 0.5 & 99.1 $\pm$ 0.3 \\ \hline
\textbf{Musical Instrument} & 70.0 $\pm$ 0.6 & 70.3 $\pm$ 0.6 & 68.3 $\pm$ 0.9 & 71.5 $\pm$ 0.2 & 72.0 $\pm$ 0.6 & \textbf{73.1 $\pm$ 0.1} \\ \hline
\textbf{Speaker Id} & 25.3 $\pm$ 0.7 & 43.5 $\pm$ 0.8 & 17.4 $\pm$ 3.4 & 31.4 $\pm$ 0.9 & 33.1 $\pm$ 0.7 & \textbf{47.7 $\pm$ 0.6} \\ \hline
\end{tabular}
\end{table*}

\subsubsection{Experimental Setup}

We have evaluated the CosCovNN on the five datasets, where all of these datasets comes with test data except IEMOCAP data. For IEMOCAP, we have calculated the accuracy based on the five fold cross validation (each fold is a session). To get a fare comparison of the performance of CosCovNN, we have used CNN with similar architecture and assessed the number of parameters for both models. Moreover, we have also compared the results with related literature Time-Domain Filterbanks (TD-filterbank) \cite{zeghidour2018learning}, SincNet \cite{ravanelli2018speaker} and LEAF \cite{zeghidour2021leaf}. Our objective is to surpass the accuracy of the CNN with our CosCovNN model while achieving accuracy levels close to those of the related literature. Additionally, visualise the filters for comparison. 

To accommodate audio signals of varying lengths, an additional layer has been incorporated at the beginning of the architecture. This layer serves to adjust the feature size to match that of a 16KHz sample rate audio signal with a duration of one second. For instance, to process audio signals with a duration of 10 seconds, a layer with a pooling size of 10 has been added. We collected the accuracy of the TD-filterbank and SincNet from the research work of Neil et al. \cite{zeghidour2018learning}, and to keep the experiment fair, we have followed the exact experimental setup from this research work. In this work, IEMOCAP and S09 dataset was not used; therefore, we have trained time-domain filterbanks, SincNet and LEAF on these datasets to collect the accuracy.

\subsubsection{Results}
The results of our experiments are presented in Table \ref{tab:cl_lit}. It is observed that CosCovNN outperforms CNN for all the tasks. Moreover, CosCovNN performs better than TD-fbanks and SincNet for all the tasks except for Acoustic Scenes and Speaker Id classification, respectively. For Acoustic Scenes classification, TD-fbanks outperforms CosCovNN. Since SincNet is explicitly designed for speaker classification, it is reasonable that it achieves better classification accuracy than CosCovNN in this regard. However, CosCovNN could not outperform LEAF in any of the tasks. This suggests that CosCovNN needs some architectural changes to surpass the best-performing model, but it can still achieve close to SOTA results.

\begin{figure*}[!t]
\centerline{\includegraphics[width=1\linewidth]{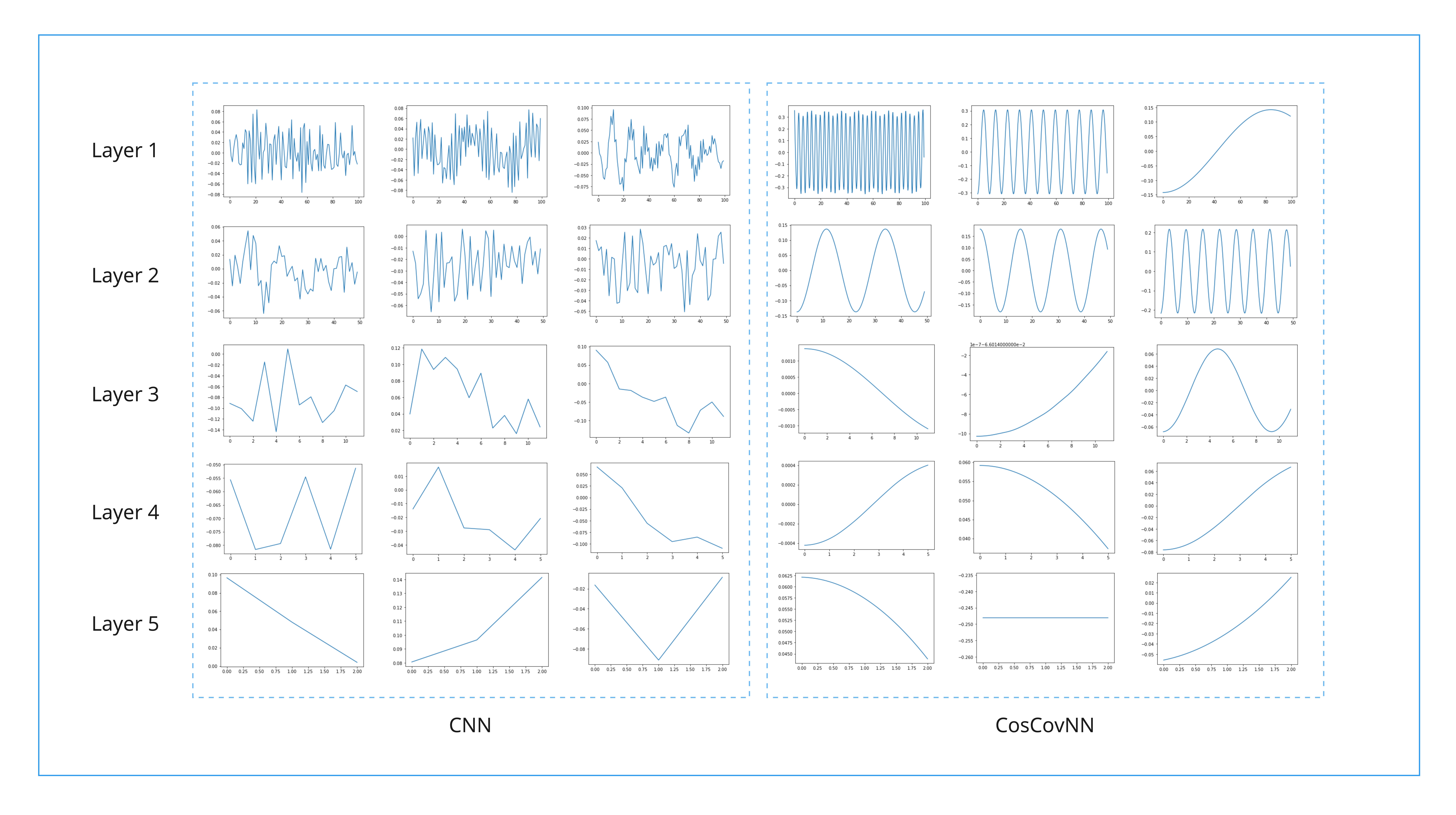}}
\caption{The figure presents a comparative visualization of the trained filter outputs from CosCovNN and traditional CNNs. On the left, the CNN filters appear irregular, reflecting their adaptation to the intricate patterns within the audio waveform through the training process. On the right, the CosCovNN filters exhibit a smoother contour, suggesting that during training, these filters have emphasised on the periodic characteristics or frequency aspects of the waveform. }
\label{fig:cvc}
\end{figure*}

Now, we can calculate the number of parameters for the CosCovNN as  $(1\times32\times2) + (32\times64\times2) + (64\times128\times2) + (128\times256\times2) + (256\times10\times2) = 91,200$ and for CNN as  $(1\times32\times100) + (32\times64\times50) + (64\times128\times12) + (128\times256\times6) + (256\times10\times3) = 4,08,192$. Notably, the CosCovNN architecture has $77.66\%$ fewer parameters than the CNN architecture, yet it outperforms the CNN. These results suggest that the cosine filters used in CosCovNN are both effective and more computationally efficient than the CNN filters in the case of audio data modelling.
The fundamental difference between CosCovNN and CNN filters is illustrated in Fig. \ref{fig:cvc}. Cosine filters are periodic, capturing critical frequency information in audio signals and are less impacted by noise. On the other hand, CNN filters try to capture the shape of the audio signal. Therefore filters are more susceptible to noise in the audio signal and less periodic.

\subsection{Comparison of VQCCM with literature}

\subsubsection{Experimental Setup}

Based on our previous experiments, we found that the CosCovNN architecture with cosine filters is more efficient than raw CNN filters. However, it is still being determined whether this approach can be used to develop a robust model that can achieve or beat SOTA performance in the literature. To address this question, we augmented the CosCovNN architecture with Memory and VQ layers to create VQCCM. We trained VQCCM with similar datasets and aimed to surpass the performance of LEAF, the best-performing model. We aim to demonstrate that the CosCovNN architecture with Memory and VQ layers can be a powerful model for audio classification tasks and achieve or surpass SOTA performance.

\subsubsection{Results}

As shown in Table. \ref{tab:cl_lit}, VQCCM has outperformed LEAF for all tasks. However, neither LEAF nor VQCCM could exceed TD-Fbanks performance in acoustic scene classification. VQCCM and LEAF achieved similar accuracy of $99.1\%$, but VQCCM has a lower standard deviation than LEAF. Furthermore, as we have only tuned VQCCM with the S09 dataset, there is an opportunity for researchers to explore and fine-tune VQCCM for each problem separately. These results demonstrate that the Cosine Convolution filter can be a solid alternative to CNN filters for raw audio classification.

\subsection{Impact of Memory and VQ size}

\subsubsection{Experimental Setup}
The VQCCM model is designed to enhance information propagation from the lower layers to the classifier layer by utilising its memory component, where the VQ layer is responsible for learning representations from specific embedding vectors. The number of vectors in the VQ layer and the size of the memory layer are two crucial factors that significantly influence the performance of VQCCM. To identify the appropriate memory size and VQ embedding numbers, we conducted separate training experiments by integrating the memory and VQ layer into the CosCovNN architecture. Initially, we planned to utilise the S09 dataset for this experimentation. However, as the performance of VQCCM and CosCovNN was found to be very similar on this dataset, the impact of the memory and VQ layer might be clear from the comparison. As a result, we expanded our experiments to include the IEMOCAP dataset. Here, we assessed the maximum accuracy based on five runs and plotted it on a graph to gain insights into its behaviour.

\subsubsection{Results}

\begin{figure*}[!t]
\centerline{\includegraphics[width=1\linewidth]{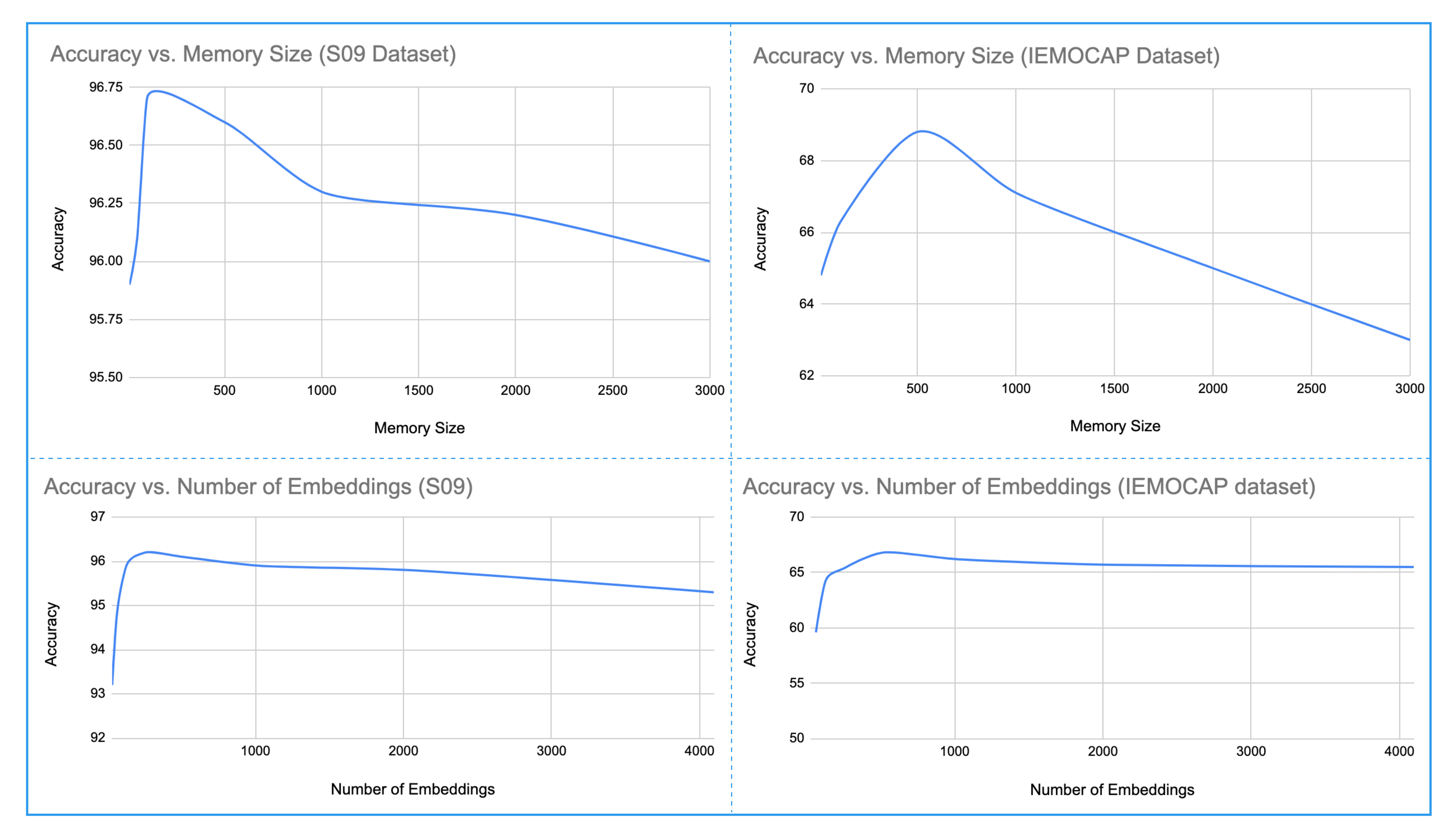}}
\caption{This figure shows the impact of the Memory and VQ layer of the VQCCM model for the S09 and IEMOCAP datasets.}
\label{fig:mvq}
\end{figure*}

The experimental outcomes are illustrated in Fig. \ref{fig:mvq}. Our results demonstrate that the model's performance can be significantly enhanced by integrating memory into it. However, it is imperative to note that an excessive increase in memory size can lead to overfitting, thereby causing a decline in performance. Conversely, setting the memory size too small, such as 10, may impair the performance of VQCCM by inadequately representing vital information from previous layers. In such cases, the multiplication of memory with the learned features in each layer negatively affects the VQCCM's performance. Prior to integrating either memory or VQ layers, our previous experiments on the S09 and IEMOCAP datasets revealed maximum accuracies of $96.4\%$ and $65.9\%$, respectively, for CosCovNN. After the inclusion of the memory layer, we achieved a maximum accuracy of $96.7 \%$ for the S09 dataset with a memory size of 100 and a maximum accuracy of $66.8 \%$ for the IEMOCAP dataset with a memory size of 500. Similarly, we obtained accuracies of $96.2 \%$ and $67.0 \%$ for the S09 and IEMOCAP datasets, respectively, with embedding sizes of 256 and 512. We observed that using a lower number of embeddings can result in underfitting while increasing the embedding size beyond a certain threshold does not cause significant degradation in performance. Unlike memory layers, overfitting the VQ layer depends on the number of feature layers, and a higher number of embeddings does not lead to overfitting.
\begin{table*}[t!]
\centering
\caption{Ablation study of the VQCCM (CosCovNN + Memory + VQ)}
\label{tab:abl}
\begin{tabular}{|l|l|l|l|l|}
\hline
\textbf{Classification Task} & \textbf{CosCovNN} & \textbf{CosCovNN + Memory} & \textbf{CosCovNN + VQ} & \textbf{CosCovNN + Memory + VQ} \\ \hline
\textbf{Speech Command} & 91.5 $\pm$ 0.2 & 94.2 $\pm$ 0.1 & 92.9 $\pm$ 0.5 & \textbf{95.6 $\pm$ 0.1} \\ \hline
\textbf{Spoken Digit} &  96.3 $\pm$ 0.1 & 96.5 $\pm$ 0.1 & 96.1 $\pm$ 0.2 & \textbf{97.1 $\pm$ 0.2} \\ \hline
\textbf{Speech Emotion}  & 63.1 $\pm$ 2.8 & 68.1 $\pm$ 0.7 & 64.1 $\pm$ 2.9 & \textbf{71.2 $\pm$ 1.4} \\ \hline
\textbf{Acoustic Scenes} & 98.3 $\pm$ 0.6 & 98.7 $\pm$ 0.3 & 98.1 $\pm$ 0.4 & \textbf{99.1 $\pm$ 0.3} \\ \hline
\textbf{Musical Instrument} & 71.5 $\pm$ 0.2 & 71.9 $\pm$ 0.1 & 70.9 $\pm$ 0.8 & \textbf{73.1 $\pm$ 0.1} \\ \hline
\textbf{Speaker Id} & 31.4 $\pm$ 0.9 & 38.2 $\pm$ 0.4 & 30.6 $\pm$ 1.6 & \textbf{47.7 $\pm$ 0.6} \\ \hline
\end{tabular}
\end{table*}

\subsection{Ablation Study for VQCCM}


In order to investigate the role of the Memory and VQ layer in the VQCCM model, we conducted an ablation study. This involved adding each component separately to the CosCovNN and observing the impact on model performance. The ablation study is equivalent to removing each component from VQCCM individually. Specifically, we performed two experiments: the first involved adding the Memory layer to the CosCovNN, resulting in a model referred to as VQCCM - VQ, and the second involved adding the VQ layer to the CosCovNN, resulting in a model referred to as VQCCM - Memory. The results of these experiments are presented in Table. \ref{tab:abl}.

Our findings indicate that adding the Memory layer to the CosCovNN results in improved performance and stability compared to CosCovNN. However, when only the VQ layer is added to the CosCovNN, improvements are not consistently observed across all experiments. While Memory is a valuable addition to CosCovNN, it is even more effective when combined with the VQ layer. The VQ layer enforces the use of a fixed number of vectors, making it difficult for the model to learn an effective representation. However, by adding the Memory layer, the model is forced to use its memory to pass important information that cannot be learned through the VQ layer alone. As a result, the presence of the VQ layer compels the model to utilise the Memory layer, leading to better results.

\section{Conclusion}
In this work, we introduce cosine filters as an innovative alternative to conventional filters in Convolutional Neural Network (CNN) models, specifically for classifying audio directly from raw waveforms. Cosine filters offer a significant advantage in computational efficiency, as each filter requires the learning of only two parameters. This is in contrast to the typically higher and more variable parameter counts associated with traditional CNN filters. We developed the CosCovNN model to implement this approach, integrating cosine filters into the CNN framework. 

Comparative analyses of CosCovNN and standard CNN architectures on Speech and Non-Speech datasets demonstrate that CosCovNN serves as an effective alternative for audio classification from raw waveforms. This study details the modifications necessary to incorporate cosine filters into CNN structures, providing practical guidelines for researchers looking to adapt existing CNN models. This adaptation not only reduces the parameter count but also has the potential to enhance performance, opening new avenues for developing CNN architectures that utilize cosine filters for audio processing.

Furthermore, we propose an advanced model, VQCCM, which builds on the CosCovNN framework by integrating Vector Quantization (VQ) and Memory layers. The classification performance of the VQCCM model was evaluated across five datasets—Speech Command, Speech Emotion, Acoustic Scenes, Musical Instrument, and Speaker Identification—where it achieved state-of-the-art results and outperformed benchmarks in certain cases. The integration of VQ and Memory layers significantly enhances the performance of CosCovNN, encouraging other researchers to incorporate these elements into various CNN architectures.

Although our findings are based on only five datasets, which presents a limitation, they highlight broader opportunities for future research. Future work could explore the applications of VQCCM in other domains such as speaker diarization, speech-to-text conversion, music mood identification, and audio event classification. Additionally, research may delve into cross-domain adaptation, integration with advanced machine learning techniques like transformers, and optimization for real-time audio processing and hardware deployment. By pursuing these directions, researchers can extend the applicability and impact of the VQCCM model, advancing the field of audio classification and contributing to the development of versatile and sophisticated audio analysis tools.

\begin{IEEEbiography}[{\includegraphics[width=1in,height=1.25in,clip,keepaspectratio]{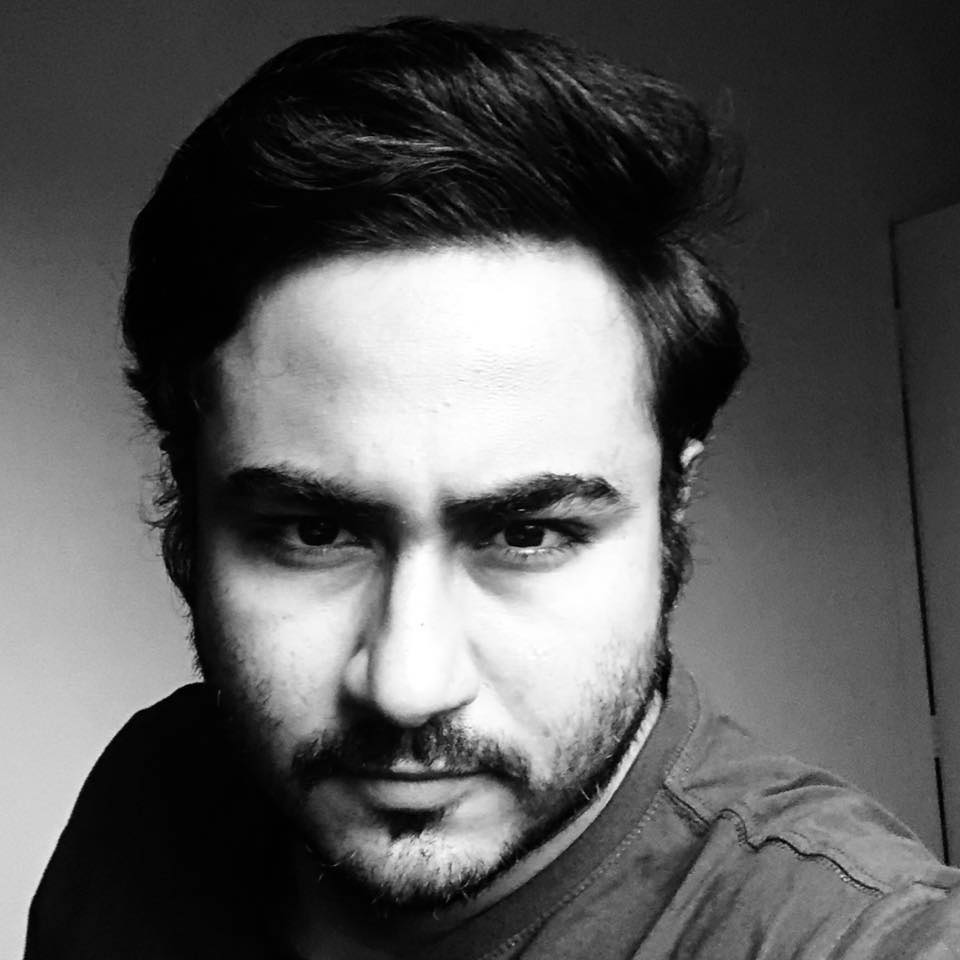}}]{Kazi Nazmul Haque} is currently pursuing his PhD at the University of Southern Queensland in Australia. Alongside his academic endeavors, he holds the position of Senior Machine Learning Engineer at Splash Music in Australia, where he specializes in the development of generative models for music. With over eight years of professional experience in machine learning, Kazi's expertise is grounded in applying machine learning techniques to address a variety of real-world challenges. His present research is primarily centered on unsupervised representation learning for audio data. Prior to embarking on his doctoral studies, Kazi earned a Master's degree in Information Technology from Jahangirnagar University in Bangladesh.
\end{IEEEbiography}

\begin{IEEEbiography}
	[{\includegraphics[width=1in,height=1.25in,clip,keepaspectratio]{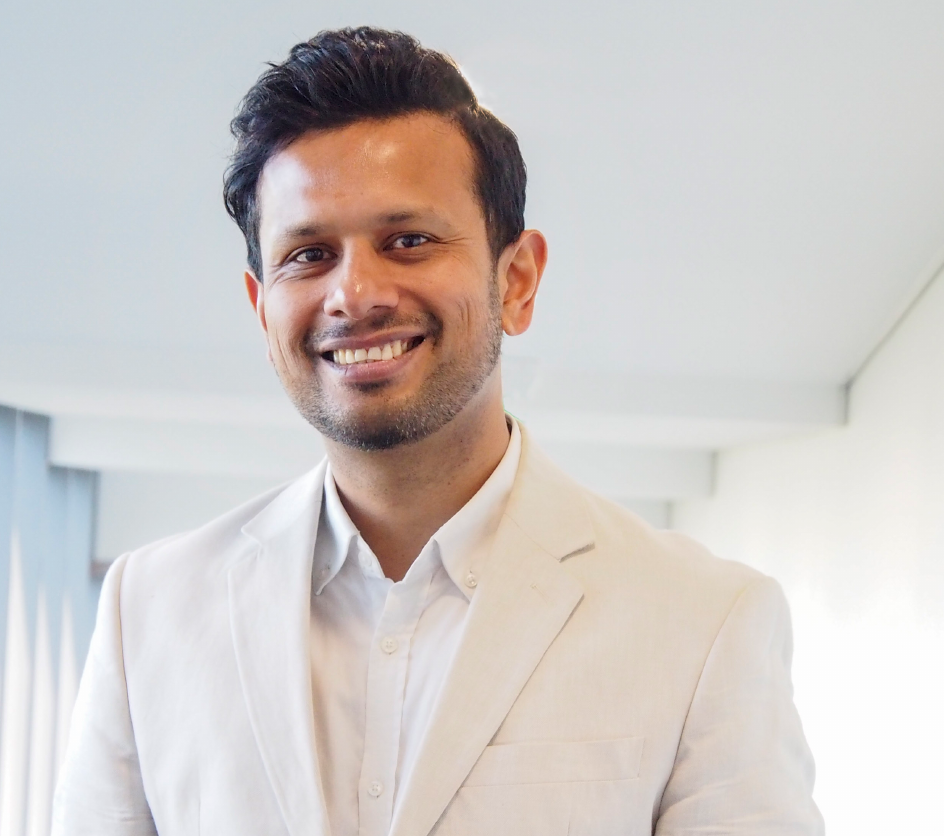}}]
	{Rajib Rana} (Member, IEEE) received his Ph.\,D.\ in Computer Science and Engineering from the University of New South Wales, Sydney, Australia, in 2011.
 
	He is an experimental computer scientist, and a Professor of Computer Science in the University of Southern Queensland. He is also the Director of the IoT Health research program at the University of Southern Queensland. He is a recipient of the prestigious Young Tall Poppy QLD Award 2018 as one of Queensland’s most outstanding scientists for achievements in the area of scientific research and communication. Rana's research work aims to capitalise on advancements in technology along with sophisticated information and data processing to better understand disease progression in chronic health conditions and develop predictive algorithms for chronic diseases, such as mental illness and cancer. His current research focus is on Unsupervised Representation Learning, Reinforcement Learning, Adversarial Machine Learning and Emotional Speech Generation. He received his B.\,Sc.\ degree in Computer Science and Engineering from Khulna University, Bangladesh, with the Prime Minister and President's Gold Medal for outstanding achievements. He received his postdoctoral training at Autonomous Systems Laboratory, CSIRO, before joining the University of Southern Queensland  in 2015.
\end{IEEEbiography}

\begin{IEEEbiography}[{\includegraphics[width=1in,height=1.25in,clip,keepaspectratio]{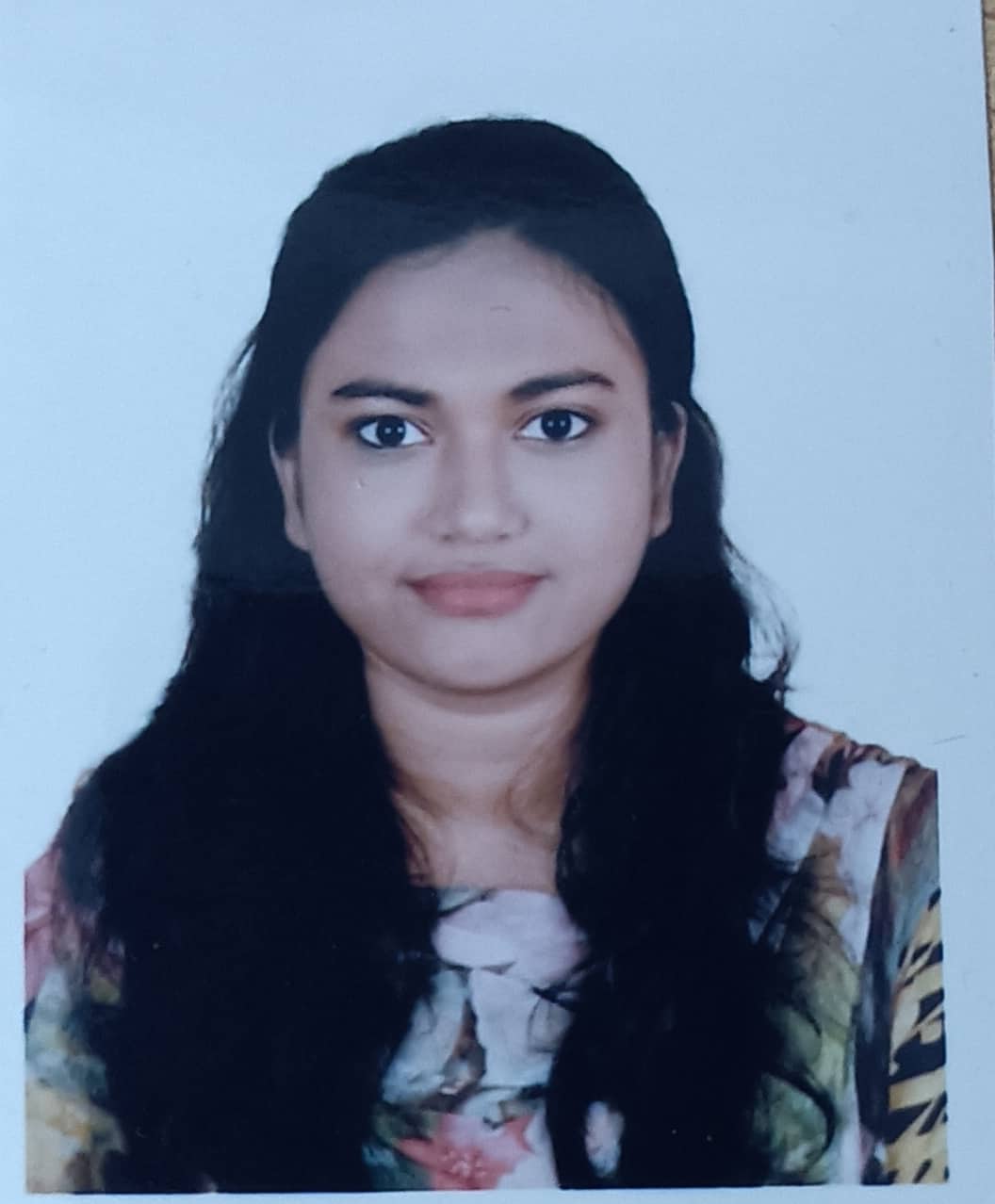}}]{Tasnim Jarin} is currently pursuing her master's degree at Jahangirnagar University,where she is actively served as a Research Assistant under the supervision of Professor Dr. Mohammad Abu Yousuf from September 2023 to September 2024.In this role, she has developed and applied advanced machine learning models to tackle complex data analysis problems, significantly enhancing model performance.
Tasnim Jarin completed her B.\,Sc.\ 
degree in Computer Science and Engineering from American International University-Bangladesh.
\end{IEEEbiography}

\begin{IEEEbiography}[{\includegraphics[width=1in,height=1.25in,clip,keepaspectratio]{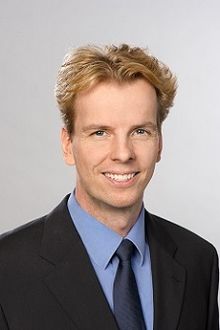}}]{Bj\"{o}rn W.\ Schuller.} (M'05-SM'15-F'18) received his diploma in 1999, his doctoral degree for his study
on Automatic Speech and Emotion Recognition in 2006, and his habilitation and Adjunct Teaching Professorship in the subject area of Signal Processing and Machine Intelligence in 2012, all
in electrical engineering and information technology from TUM in Munich/Germany. He is Professor of Artificial Intelligence in the Department of Computing at the Imperial College London/UK, where he heads GLAM –- the Group on Language, Audio \& Music, Full Professor and head of the Chair of Embedded Intelligence for Health Care and Wellbeing at the University of Augsburg/Germany, and founding CEO/CSO of audEERING. He was previously full professor and head of the Chair of Complex and Intelligent Systems at the University of Passau/Germany.

Professor Schuller is Fellow of the IEEE, Golden Core Member of the IEEE Computer Society, Fellow of the ISCA, Senior Member of the ACM, President-emeritus of the Association for the Advancement of Affective Computing (AAAC), and was elected member of the IEEE Speech and Language Processing Technical Committee. He (co-)authored 5 books and more than 900 publications in peer-reviewed books, journals, and conference proceedings leading to more than overall 30\,000 citations (h-index = 82). Schuller is Field Chief Editor of Frontiers in Digital Health, former Editor in Chief of the IEEE Transactions on Affective Computing, and was general chair of ACII 2019, co-Program Chair of Interspeech 2019 and ICMI 2019, repeated Area Chair of ICASSP, next to a multitude of further Associate and Guest Editor roles and functions in Technical and Organisational Committees.
\end{IEEEbiography}

\end{document}